# An Iterative Algorithm for Battery-Aware Task Scheduling on Portable Computing Platforms*

Jawad Khan and Ranga Vemuri

*ECECS Department, University of Cincinnati, Cincinnati, Ohio 45221-0030, USA*

*{jkhan,ranga}@ececs.uc.edu*

**Abstract**

*In this work we consider battery powered portable systems which either have Field Programmable Gate Arrays (FPGA) or voltage and frequency scalable processors as their main processing element. An application is modeled in the form of a precedence task graph at a coarse level of granularity. We assume that for each task in the task graph several unique design-points are available which correspond to different hardware implementations for FPGAs and different voltage-frequency combinations for processors. It is assumed that performance and total power consumption estimates for each design-point are available for any given portable platfrom, including the peripheral components such as memory and display power usage. We present an iterative heuristic algorithm which finds a sequence of tasks along with an appropriate design-point for each task, such that a deadline is met and the amount of battery energy used is as small as possible. A detailed illustrative example along with a case study of a real-world application of a robotic arm controller which demonstrates the usefulness of our algorithm is also presented.*

## 1. Introduction

Battery powered portable systems have finite amount of battery energy available and therefore battery lifetime maximization is one of the most important design goals for such systems. In this paper we present an iterative heuristic algorithm based on the battery discharge characteristics. Our goal is to meet a desired deadline and save as much battery energy as possible.

**Target Hardware Architecture:** The algorithm described in this work is applicable to any embedded platform, although the methods for changing the energy consumption of a task vary depending upon which processing element is used. In a processor based embedded system power-performance trade-offs can be achieved by voltage and clock scaling. It is assumed that several discrete voltage and frequency combinations are available. If the embedded platform has an FPGA as the main processing element then it is assumed that several different hardware implementations are available which can be downloaded in the form of bitstreams. It is assumed that for each design-point, performance and total power cosumption estimates are available for any given portable platfrom, including the peripheral components such as memory and display power usage. Further, it is also assumed that intertask communication occurs via shared memory and the energy cost and latency of the memory transfers are a part of the execution time and energy costs of the task under consideration.

**Application Specification:** The application is described as a directed acyclic task graph (DAG) $G(V,E)$. The vertices (nodes) of the graph are tasks which are to be executed on the portable platform and each task $v \in V$ has several different implementation options available called design-points. Associated with each task $i$ and its design-point $j$ is its execution time $E_{ij}$ and current consumption $I_{ij}$. The current consumption of a task is assumed to be measured as the average total current consumption of the portable platform which is the cumulative current consumption of all subsystems being used in the portable platform. The edges $E$ describe data and control dependence between different tasks of the task graph. In the rest of the paper we will use $n = |V|$, $e = |E|$ and there are $m$ design points available for each task. There is a deadline $d$ associated with the task graph before which all the tasks must be completed.

**Problem Description:** Given a DAG $G(V,E)$, a set of design-points for each task, execution time, current usage estimate for each design-point and a desired deadline for the completion of the task graph, determine a valid schedule which does not violate the control and data dependencies of the tasks in the task graph and also find a mapping of each task to a suitable design-point such that the deadline for the entire task-graph is met and the battery energy used is as small as possible. Our algorithm finds an initial sequence and then assigns design-points to the tasks. The suitability of the task sequence and the design-point selection is judged by using a battery model due to [2] which will be discussed in Section 3. The chosen task sequence is modified using a heuristic approach and design-point selection is performed again to improve the quality of the solution. The algorithm is described in detail in Section 4. We tested the algorithm using different task-graphs and design-points and the results are discussed in Section 5.

## 2. Related Work

In processor based embedded computing platforms dynamic voltage and frequency scaling has been proven to be extremely effective for low power execution of tasks [4][5][6][8]. In battery-powered embedded systems the energy source is non-linear therefore, the existing voltage scaling techniques are not directly applicable to these systems. Luo and Jha studied static task scheduling for battery powered multiprocessor environments [5]. They used a battery model which was based on Peukert's law and an empirical model due to Pedram [6]. Rakhmatov et al developed an algorithm for battery-aware task scheduling using dynamic programming [1] along with its other variants. Chowdhury et al [7] proposed a simplified heuristic which reduced the voltage level of the tasks as much as possible starting from the last task in the schedule. In contrast to these works, our work provides a way to simultaneously solve the task sequencing and design-point assignment in an iterative fashion. In any given iteration a valid schedule and assignment is available which can be used. If the user wishes then the solution quality is improved over subsequent iterations. Further, compared to our algorithm, it is not easy to implement a Simulated Annealing or Linear Program Forumulation based algorithms on an embedded computing platform which has inherent limitations on memory and battery capacity.

## 3. Battery Characteristics and Motivation

*Rated Capacity* of a battery is defined as the capacity of the battery (in *mAh*) under a nominal constant current discharge and is reported by the manufacturer. It is observed that higher rates of discharge tend to reduce the rated capacity significantly (rate capacity effect) and reducing discharge rates between heavy discharge periods allows the battery to regain some of its lost capacity (recovery effect) [3]. Rakhmatov et al. [2] developed a variable load analytical model based on the laws of chemical kinetics, which takes into account both the rate capacity effect and the recovery effect. Equation 1 describes the battery model.

$$\sigma = \sum_{k=0}^{n-1} I_k \left( \Delta_k + 2 \sum_{m=1}^{10} \frac{e^{-\beta^2 m^2 (T - t_k - \Delta_k)} - e^{-\beta^2 m^2 (T - t_k)}}{\beta^2 m^2} \right) \quad (1)$$

The value of $\sigma$ gives the amount of charge lost by time $T$, which is the length of a current discharge profile having $n$ distinct discharge intervals. $I_k$ is the current drawn from the battery in the $k^{th}$ discharge interval, where $t_k$ is the start time of the $k$th discharge interval and $\Delta_k$ is the duration of this interval. The battery lifetime is estimated by evaluating Equation 1 for increasing values of $T$ and stopping where $\sigma \cong \alpha$: At this point the value to $T$ is taken as the battery lifetime. Equation 1 is used as the battery-aware cost function to be minimized. We have chosen to use this battery model because of its high accuracy and low computational complexity. It was shown in [1] that for a set of n tasks if dependencies are ignored and the value of $\alpha$ is assumed to be sufficiently large then sequencing tasks in the non increasing order of their currents is the best and sequencing the tasks in the non decreasing order of their currents is the worst. This property is also important for task-graphs where dependencies are present because it provides the lower and upper bounds on the value of cost function given in Equation 1. The authors in [7] also proved that given a pair of two identical tasks in the profile and a delay slack to be utilized by down scaling, it is always better to use the slack on the later task than on earlier task. We use the above two properties along with the observation that tasks which have lower overall average energy consumption should be given priority for voltage down scaling.

## 4. Battery-Aware Task Sequencing and Design-Point Assignment

Some important definitions are presented below first:

**Execution Time matrix (D)** is an (*n* x *m*) matrix where $D_{i,j}$ gives the execution time of task *i* using design-point *j*; for each task *i* the execution times of the design-points are stored in ascending order of magnitude.

**Current matrix (I)** is an (*n* x *m*) matrix where $I_{i,j}$ gives current of task *i* using design-point *j*; for each task *i* the currents of the design-points are stored in descending order of magnitude.

**Design-Point Selection matrix (S)** is an (*n* x *m*) matrix where $S_{i,j}$ is 1 if task *i* is assigned to design-point *j*. S is initialized such that $S_{i,j}$ ={1 if *j* = *m*, 0 otherwise}.

**Energy Vector E:** is a row vector where each element specifies a task and the tasks are stored in increasing order of their average energies.

**Slack Ratio (*SR*)** of a design-point is defined as the ratio of the amount of slack left to the deadline, if that design-point is chosen for execution. If *t* is the execution time of a design-point and *d* the deadline of the task graph then formally *SR* is defined as: $SR = (d - t)/(d)$. Similarly, if *SR* is to be calculated for several design-points chosen, then *t* would be defined as the sum of the execution times of all the design-points chosen. *SR* gives an indication of how much slack is left which needs to be utilized. It is beneficial to use as much slack as possible. Therefore, a smaller value of *SR* is better.

**Current Ratio (*CR*)** of a design-point is defined as $CR = (I - I_{min})/(I_{max} - I_{min})$, where *I* is the average current used by the design-point and $I_{max}$ and $I_{min}$ are the maximum and minimum currents among all the design-points of all tasks. *CR* is normalized to be between 0 and 1. *CR* gives an indication of relative current of a particular design-point when compared to all the other design-points. A smaller value of *CR* is better.

**Energy Ratio (*ENR*)** of a task sequence is defined as $ENR = (E_n - E_{min})/(E_{max} - E_{min})$, where $E_n$ is the total average energy used by the chosen design-points for all the tasks. Energy ratio is low if a set of design-points uses lower overall average energy. Its value is between 0 and 1.

$$E_n = \sum_{i=1}^{n-1} I_{i,c} \times V_{i,c} \times D_{i,c}$$

$I_{i,c}$ and $V_{i,c}$ are the current and the voltage of the design-point *c* chosen for task *i*, respectively. and $D_{i,c}$ is its corresponding execution time. $E_{min}$ and $E_{max}$ are the energies of tasks sequences if all the lowest and highest power design-points are used for all tasks, respectively.

$$E_{min} = \sum_{i=1}^{n-1} I_{i,min} \times D_{i,min} \qquad E_{max} = \sum_{i=1}^{n-1} I_{i,max} \times D_{i,max}$$

**Current Increase Fraction (*CIF*)** of a task sequence is a measure of non-decreasing trends in the current discharge profile. The lower its value, the less number of increasing current transitions are there in the discharge profile. We define *CIF* as follows where *c* denotes the chosen design point for any task *k*:

$$CIF = \sum_{k=2}^{n} J_k/n - 1 \text{ where } J_k = \begin{cases} 1 & \text{if } I_{k-1, c} < I_{k, c} \\ 0 & \text{Otherwise} \end{cases} \quad k = 2...n$$

**Design-Point Fraction (*DPF*):** If there are a total of *m* design points available for each task then *DPF* is the fraction of total design-points assigned to any single design-point *k* for all *n* tasks. *DPF* is a measure of the number of different design-points being used in a particular assignment. Equation 2 and Equation 3 formally define *DPF* where *k* denotes the design point under consideration where *x* is the number of free nodes:

$$DPF = \sum_{k=1}^{m} (m-k) \times f \times F_k \quad (2)$$

$$f = 1/(m-1) \quad F_k = \sum_{i=1}^{x} S_{i,k}/(x) \quad \forall k \in 1...m \quad (3)$$

As can be seen from the definition, the use of higher powered design-points is penalized the most. The penalization decreases as the lower powered design points are used and is zero for the lowest powered design-points.

**Suitability of a Design-Point (*B*)** is a measure of how suitable a particular design-point is for achieving the minimum battery capacity usage goal and the suitability of a design-point is defined as $B = SR + CR + ENR + CIF + DPF$

### 4.1 Algorithm Description

During each iteration of the algorithm a valid schedule is created and a design-point assignment is chosen. In subsequent iterations the solution is improved such that the amount of battery capacity used is less than the previous iteration. The algorithm terminates once the solution is not improved during two consecutive iterations. We start choosing design-points from the last task and work our way up to the first task. During the selection of the design-point for any particular task the suitability of the each design-point *(B)* is calculated. The design-point which has the lowest value of *B* is chosen and the task is fixed to that design-point. Each task can be in three different states of design-point allocation: free, tagged and fixed. A task is tagged when we are evaluating the suitability of one of its design-points. When all of its design-points are evaluated and the best one is selected we set the state of the task to be fixed. Tasks which are neither tagged nor fixed are called free tasks. We use an heuristic approach for searching the design-space which involves the use of a window function (Explained later) until all design-points are considered. Once all tasks are assigned to design-points the battery capacity used is calculated and we sequence the tasks according to a weight assignment which is based on the current consumption of the design-points assigned to different tasks in this iteration (Explained in Section 4.). This new sequence is then used for the next iteration.

*BatteryAwareSQNDPAllocation* is the top level algorithm for task sequencing and design-point assignment and is given in Figure 1. *MinBCost* contains the minimum battery cost for any given iteration and is initialized to infinity. *PrevIterCost* contains the cost of the previous iteration for comparison pur-

```
BatteryAwareSQNDPAllocation
Begin
MinBCost = infinity, PrevIterCost = infinity
L=SequenceDecEnergy(D,I)
success = TRUE
while (success)
    {MinBCost,S}=EvaluateWindows(L,E,I,D)
    Ltemp = FindWeightedSequence(S,I,D)
    TempCost = CalculateBatteryCost(Ltemp,S,I,D)
    if TempCost < MinBCost then
        MinBCost = TempCost
        success = TRUE;
    end if
    if MinBCost >= PrevIterCost then
        success = FALSE
    else
        PrevIterCost = MinBCost
    end if
    L = Ltemp
end while
```

```
{MinBCost,S}=EvaluateWindows(L,E,I,D)
Wflag = TRUE
WindowStart = m-1
while (Wflag)
    if (d < C_T(WindowStart)) then
        WindowStart = WindowStart - 1
        Wflag = TRUE
        if (WindowStart == 1) AND (d < C_T(WindowStart)) then
            Exit with error // The deadline cannot be met
        end if
    else  Wflag = FALSE   end if
end while
while ( windowstart >= 1 )
    //Use WindowStart .... m columns of S, I, D for this iteration
    Stemp = ChooseDesignPoints(L,E,I,D,WindowStart,d)
    TempCost = CalculateBatteryCost(L,Stemp,I,D)
    if TempCost < MinBCost then
        MinBCost = TempCost;
        S = Stemp
    end if
    WindowStart = WindowStart - 1
end while
return{MinBCost,S}
```

```
{S} = ChooseDesignPoints(L,E,I,D,WindowStart,d)
Initialize S, Free all tasks in E
Tsum = D(n,m)
Ttemp = 0
Mark the task n in E to be fixed
S(n,m) = 1  //Assign nth task to the lowest power design-point
for i = n-1 downto 1 do
  for j = m downto WindowStart do
    S(i,j) = 1
    Ttemp = Tsum + D(i,j)
    Set task i to be tagged in S and fix it in E
    SR = (d - Ttemp)/d
    CR = (I(i,j) - Imin)/(Imax - Imin)
    {ENR,CIF,DPF} = CalculateDPF(E,S,I,D,WindowStart,i,d)
    B(i,j) = SR + CR + ENR + CIF + DPF
    S(i,j) = 0, Ttemp = 0
  end for
  find B(i,k) the minimum value of B for task i
  S(i,k) = 1
  Set task i to be fixed in S and E
  Tsum = Tsum + D(i,k)
end for
return{S}
```

**Figure 1. Algorithms for Battery-Aware Task Sequencing and Design-Point Allocation**

```
{ENR,CIF,DPF} = CalculateDPF(E,S,T,D,WindowStart,i,d)
Etemp=E, Stemp=S
flag = FALSE
T_e = CalculateExecutionTime(Stemp,D)
while (T_e > d)
  Choose the first free task q in Etemp
  r = the row at which task q is located in Stemp
  if no free task found then
    DPF = infinity
    {CIF,ENR} = CalculateFactors(Stemp,I,D)
    return {ENR,CIF,DPF}
  else
    for p = m downto WindowStart do
      if (p = WindowStart+1) then Fix node q in Etemp end if
      if Stemp(r,p) == 1 then
        Stemp(r,p) = 0, Stemp(r,p-1) = 1, break
      end if
    end for
  end if
  T_e = CalculateExecutionTime(Stemp,D)
end while
DPF = 0, ufac = m - WindowStart, factor = 1/ufac
for w = 1 to (m-WindowStart) do
  DPF = DPF + ufac × factor × Σ_{y=1}^{i-1} (Stemp_{y,w})/(i-1)
  ufac = ufac - 1
end for
if this is the last free task then
  DPF = (d - Te)/d
end if
{CIF,ENR} = CalculateFactors(Stemp,I,D)
return{ENR,CIF,DPF}
```

```
{CIF,ENR}=CaculateFactors(S,I,D)
Ttemp = S * D // where * denotes element by element
Itemp = S * I // multiplication of two matrices of same dimensions
Time = Column by Column sum of matrix Ttemp
// Time and Current are column vectors
Current = Column by Column sum of matrix Itemp
CIF = 0
for x = 2,..,n do
  if Current(x-1) < Current(x) then
    CIF = CIF + 1
  end if
end for
CIF = CIF /( total number of tasks - 1)
E_n = Current * Time
Energy = sum of all rows of E_n
ENR = (Energy - Emin)/(Emax - Emin)
return{CIF,ENR}
```

**Figure 2. Algorithms for various factor calculations**

poses and is also initialized to infinity. We use a modified list based scheduling algorithm for generating task sequences for a particular task graph. At the start of the algorithm each task is assigned a weight which is equal to the average energy of all design-points available for it. The tasks which have a larger weight are scheduled earlier than other tasks in the ready list, which is a list of all nodes which have all their predecessors scheduled. The scheduled list is called L(1..n) and is generated by the algorithm called *SequenceDecEnergy(D,I)*. Each task is assigned a unique time-step and all the $n$ tasks in this list are scheduled to be executed sequentially. The actual algorithm is omitted due relatively straight-forward nature. *success* is a flag which is set whenever there is a better solution found in the present iteration. If the solution does not improve over two consecutive iterations the algorithm terminates with the best solution found.

**Window Function:** In each iteration a window dictates how many design-points for each task are to be considered for allocation. For example, consider five tasks and four design-points as shown in Figure 3. The three different windows basically mask out all the columns which are beyond the width of the window and only the design-points within the window are evaluated. *EvaluateWindows(L,E,I,D)* initially tries to find a valid start width *(WindowStart)* for the window. $C_T(k)$ is the execution time if all design-points belonging to column $k$ are chosen. The algorithm checks whether it would be possible to meet the deadline by executing even the highest power design-points or not: if $d < C_T(1)$ then the deadline cannot be met and the algorithm exits with an error. Otherwise the window width is incrementally increased until all $m$ design-points are evaluated. The design-point allocation *(S)* which results in the least amount of battery cost *(MinBCost)* is then returned.

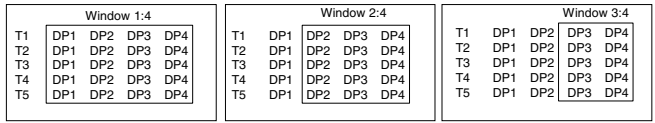

**Figure 3. Various windows**

**Choosing Design Points:** In *ChooseDesignPoints()* we first initialize S and set the state of all nodes in E to be free. Recall that E is the Energy Vector. We will use E as a priority function while evaluating DPF. We start from the last task in the sequence and fix it to the lowest power design-point and move up towards the first task. *Tsum* keeps track of the sum of the execution times of the tasks fixed so far. *Ttemp* is used to keep track of the sum of execution time of the tagged tasks as well as the sum of the execution time of the fixed tasks. We evaluate the suitability *B* of each design point $j$ of each task one by one. For any given task $i$ the design-point $k$ which has the minimum value of *B* is chosen and the task $i$ is fixed to design point $k$ and the value of *Tsum* is updated to reflect the chosen design-point for task $i$

**Calculation of DPF, CIF and ENR:** The algorithms given in Figure 2 are used to calculate DPF, CIF and ENR. Initially *CalculateDPF()* is called from *ChooseDesignPoints()* Then it calls *CaculateFactors()* to calculate *ENR* and *CIF*. Copies of *S* and *E* are made as *Stemp* and *Etemp*.

At any given point during the execution there will be some tasks in *S* which have been fixed, tagged and free. Similarly, each task in *E(Etemp)* has two states: fixed and free. A task is fixed in *E* if it is fixed or tagged in *S* and also when the highest powered design-point is chosen for it. If the deadline is not being met by choosing the lowest power consuming design-points of all free tasks in E(*Etemp*), moving the first free task in E(*Etemp*) from lower to a higher power consuming design-point decreases the execution time in the hope that the deadline will be met with the least increase in the overall energy consumption. For example, consider the tasks and their design-points shown in Figure 4. Here tasks T5 and T4 have been fixed to DP4 and DP1, respectively. T3 is the tagged task and we are calculating the *DPF* of DP2. Figure 4-a shows the initial condition at the start of calculation of DPF. Notice tasks T1 and T2 are both initially assigned to DP4. Suppose, that it is found that this assignment does not meet the deadline. Now we wish to use a higher powered DP such that the deadline is met. We look at Energy vector E and find that tasks 3,4,5 are fixed. The first free task in E is T1. Therefore, it is assigned the next higher powered design-point, DP3 (Figure 4-b). Suppose, that this assignment also, does not meet the deadline. The assignment of T1 is moved up to DP2 (Figure 4-c). Now, let us assume that the assignment meets the deadline. Therefore, no further moves are necessary. From Equation 2 and Equation 3 $f = 1/3$ and the number of free nodes $x$ is 2. $F_4 = 1/2$, $F_3 = 0$, $F_2 = 1/2$, $F_1 = 0$, therefore $DPF = 1/3$ for task T3, DP2.

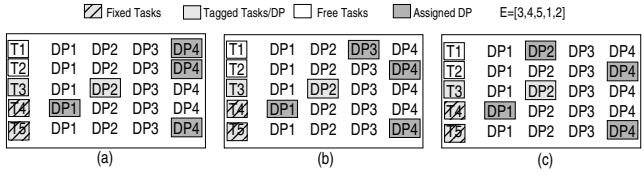

**Figure 4. DPF Calculation**

This process is repeated until either the deadline is met or there are no more free tasks available in $E(Etemp)$. If no free tasks are found in $E(Etemp)$, the value of $DPF$ is set to infinity to indicate that choosing the corresponding tagged designpoint would result in a deadline violation. Finally, if we are considering the last task we set $DPF$ equal to the slack ratio so that more emphasis is given to decreasing the slack. **CaculateFactors()** is called at the end to calculate the *CIF* and *ENR* based upon the design-point allocation decisions made in the calculation of *DPF*. CIF basically tries to capture the increasing current profile in any design-point assignment. *ENR* gives the overall energy consumption of the design-point assignment.

**Calculation of the Weighted Sequence:** The main subroutine called *BatteryAwareSQNDPAllocation* calls *FindWeightedSequence()* after different windows are evaluated. We try to improve the sequence by assigning weights to each task $v$ according to Equation 4 where $G_v$ is the sub-graph rooted at the node $v$. A list based scheduling method similar to *SequenceDecEnergy* is used for this algorithm as well but with modified weights.

$$\forall v \in G \quad w(v) = \sum_{\forall v \in G_v} I_v \quad (4)$$

The battery cost is calculated using Equation 1 for the task sequence generated and the design-points selected. This is done using the function called *CalculateBatteryCost()*. If there is no improvement in the battery cost in two successive iterations, the algorithm terminates.

## 4.2 Illustrative Example

We demonstrate the working of the algorithm with the help of the task graph (G3) shown in Table 1, which has 15 tasks and five different design points. For G3 the task durations were proportional to the worst case execution of the tasks and were made inversely proportional to the scaling factor with respect to voltage of DP1 ($V_1$) and task currents for differnent designpoints were made directly proportional to the cube of the scaling factor with respect to $V_1$. The scaling factors used for the five design points with respect to $V_1$ were as follows: 1, 0.85, 0.68, 0.51, 0.33. This task graph corresponds to a class of task graphs called fork-join, such task graphs have been used in multiprocessor scheduling research to model the structure of commonly encountered parallel algorithms [9]. The dependency constraints are listed under the column called "Parents". We let the deadline to be 230 minutes, $\beta = 0.273$ and executed our algorithm on G3, also we assumed that the amount of battery capacity available ($\alpha$) was sufficiently large to accommodate the requirements of different tasks.

Despite the unually large values for the task durations used in this example the algorithm is equally applicable to any chosen time scale. Table 2 shows the task sequences generated for the four iterations of the algorithm along with the designpoints (DP) assigned for each sequence. A "w" after a sequence number indicates the weighted sequence calculated for the corresponding iteration.

**Table 1. Data for example task graph G3**

| Tasks | Design Point 1 | | Design Point 2 | | Design Point 3 | | Design Point 4 | | Design Point 5 | | Parents |
|---|---|---|---|---|---|---|---|---|---|---|---|
| | I mA | D min | I mA | D min | I mA | D min | I mA | D min | I mA | D min | |
| T1 | 917 | 7.3 | 563 | 11.2 | 288 | 15.0 | 122 | 18.7 | 33 | 22.0 | - |
| T2 | 519 | 11.2 | 319 | 17.3 | 163 | 23.1 | 69 | 28.9 | 19 | 34.0 | T1 |
| T3 | 611 | 5.9 | 375 | 9.2 | 192 | 12.2 | 81 | 15.3 | 22 | 18.0 | T1 |
| T4 | 938 | 5.3 | 576 | 8.2 | 295 | 10.9 | 124 | 13.6 | 34 | 16.0 | T1 |
| T5 | 781 | 4.0 | 480 | 6.1 | 246 | 8.2 | 104 | 10.2 | 28 | 12.0 | T1 |
| T6 | 800 | 4.6 | 491 | 7.1 | 252 | 9.5 | 106 | 11.9 | 29 | 14.0 | T2,T3 |
| T7 | 720 | 7.3 | 442 | 11.2 | 226 | 15.0 | 96 | 18.7 | 26 | 22.0 | T4,T5 |
| T8 | 600 | 5.3 | 368 | 8.2 | 189 | 10.9 | 80 | 13.6 | 22 | 16.0 | T6,T7 |
| T9 | 650 | 4.6 | 399 | 7.1 | 204 | 9.5 | 86 | 11.9 | 23 | 14.0 | T8 |
| T10 | 710 | 5.9 | 436 | 9.2 | 223 | 12.2 | 94 | 15.3 | 26 | 18.0 | T8 |
| T11 | 500 | 6.6 | 307 | 10.2 | 157 | 13.6 | 66 | 17.0 | 18 | 20.0 | T9 |
| T12 | 510 | 4.6 | 313 | 7.1 | 160 | 9.5 | 68 | 11.9 | 18 | 14.0 | T10 |
| T13 | 700 | 4.0 | 430 | 6.1 | 220 | 8.2 | 93 | 10.2 | 25 | 12.0 | T9 |
| T14 | 400 | 5.3 | 246 | 8.2 | 126 | 10.9 | 53 | 13.6 | 14 | 16.0 | T11,T12T13 |
| T15 | 380 | 3.3 | 233 | 5.1 | 119 | 6.8 | 50 | 8.5 | 14 | 10.0 | T14 |

**Table 2. : Task Sequences of G3 for different iterations**

| Iter | Seq No | Task Sequences |
|---|---|---|
| 1 | S1 | T1,T4,T5,T7,T3,T2,T6,T8,T10,T9,T13,T12,T11,T14,T15 |
| | DP | P5,P5,P5,P4,P4,P4,P4,P4,P4,P4,P4,P4,P4,P4,P5, |
| | S1w | T1,T3,T2,T4,T5,T6,T7,T8,T10,T9,T13,T12,T11,T14,T15 |
| 2 | S2 | T1,T3,T2,T4,T5,T6,T7,T8,T10,T9,T13,T12,T11,T14,T15 |
| | DP | P5,P1,P2,P5,P5,P5,P5,P5,P5,P5,P5,P5,P5,P5,P5 |
| | S2w | T1,T3,T2,T4,T5,T6,T7,T8,T9,T10,T13,T11,T12,T14,T15 |
| 3 | S3 | T1,T3,T2,T4,T5,T6,T7,T8,T9,T10,T13,T11,T12,T14,T15 |
| | DP | P5,P5,P1,P5,P5,P4,P5,P4,,P5,P5,P5,P5,P5 |
| | S3w | T1,T2,T4,T5,T7,T3,T6,T8,T9,T10,T13,T11,T12,T14,T15 |
| 4 | S4 | T1,T2,T4,T5,T7,T3,T6,T8,T9,T10,T13,T11,T12,T14,T15 |
| | DP | P5,P1,P5,P5,P4,P5,P5,P5,P4,,P5,P5,P5,P5,P5 |
| | S4w | T1,T2,T4,T5,T7,T3,T6,T8,T9,T10,T13,T11,T12,T14,T15 |

Table 3 gives the values of battery capacity used σ given in milli-Ampere-minutes along with the duration of the task sequence Δ given in minutes, for each sequence and its weigthed counterpart for the four iterations of the algorithm execution. The column marked "Win 1:5" contains the results for the battery capacity and the execution time when the algorithm was allowed to consider all five design points for all tasks. The data for each subsequent window where the number of allowed design points for consideration were decreased by one is also given in Table 3. The column called "Min σ" contains the minimum value of battery capacity chosen from among the four different windows evaluated during an iteration. The last column gives the execution time corresponding to the window chosen which uses the minimum battery capacity. The battery capacity after the first iteration is 16353 $mAmin$ which decreases after each subsequent iteration until iteration 3. For iteration 4 there is no improvement in the value of the battery capacity used and the algorithm terminates. Also notice that for each iteration a valid schedule is generated which satisfies the deadline.

**Table 3. : Algorithm execution data for different iterations for G3**

| Seq No | Win 1:5 | | Win 2:5 | | Win 3:5 | | Win 4:5 | | Min σ | Δ |
|---|---|---|---|---|---|---|---|---|---|---|
| | σ | Δ | σ | Δ | σ | Δ | σ | Δ | | |
| S1 | 17169 | 229.8 | 17837 | 228.4 | 17038 | 227.1 | 16353 | 228.3 | 16353 | 228.3 |
| S1w | - | - | - | - | - | - | - | - | 16353 | 228.3 |
| S2 | 14725 | 229.2 | 16126 | 229.2 | 15929 | 229 | 16235 | 229.2 | 14725 | 229.2 |
| S2w | - | - | - | - | - | - | - | - | 14725 | 229.2 |
| S3 | 13737 | 229.8 | 16033 | 229.2 | 16061 | 229.8 | 16677 | 228.9 | 13737 | 229.8 |
| S3w | - | - | - | - | - | - | - | - | 13737 | 229.8 |
| S4 | 13737 | 229.8 | 15866 | 229.3 | 16240 | 229.2 | - | - | 13737 | 229.8 |
| S4w | - | - | - | - | - | - | - | - | 13737 | 229.8 |

## 5. Case Study

We present a real-world application of a robotic arm controller implemented on a voltage scalable processor as described in [10][1] to demonstrate the usefulness of this algorithm. The task-graph of this application (called G2) is shown in Figure 5 along with its various design-points. For G2 the task durations were proportional to the worst case execution of the tasks and were made inversely proportional to the scaling factor with respect to voltage of DP4 ($V_4$) and task currents for differnent design-points were made directly proportional to the cube of the scaling factor with respect to $V_4$. The scaling factors used for the four design points with respect to $V_4$ were as follows: 2.5, 1.66, 1.25, 1. The battery capacities used for three different deadlines (55, 75, 95 minutes) are shown in Table 4.

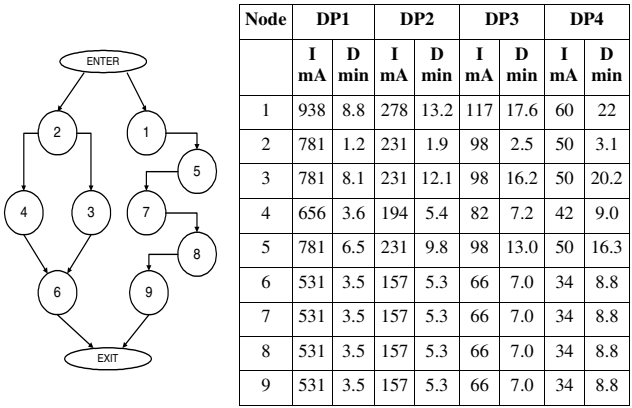

| Node | DP1 | | DP2 | | DP3 | | DP4 | |
|---|---|---|---|---|---|---|---|---|
| | I mA | D min | I mA | D min | I mA | D min | I mA | D min |
| 1 | 938 | 8.8 | 278 | 13.2 | 117 | 17.6 | 60 | 22 |
| 2 | 781 | 1.2 | 231 | 1.9 | 98 | 2.5 | 50 | 3.1 |
| 3 | 781 | 8.1 | 231 | 12.1 | 98 | 16.2 | 50 | 20.2 |
| 4 | 656 | 3.6 | 194 | 5.4 | 82 | 7.2 | 42 | 9.0 |
| 5 | 781 | 6.5 | 231 | 9.8 | 98 | 13.0 | 50 | 16.3 |
| 6 | 531 | 3.5 | 157 | 5.3 | 66 | 7.0 | 34 | 8.8 |
| 7 | 531 | 3.5 | 157 | 5.3 | 66 | 7.0 | 34 | 8.8 |
| 8 | 531 | 3.5 | 157 | 5.3 | 66 | 7.0 | 34 | 8.8 |
| 9 | 531 | 3.5 | 157 | 5.3 | 66 | 7.0 | 34 | 8.8 |

**Figure 5. Task Graph G2 and Design-Point Data**

**Comparison with an Approach in [1]:** We compared the results from our algorithm to a method in [1] where the design points were chosen using a dynamic program such that the total energy used is minimized and a given deadline is met. In the algorithm given in [1], after the design-point allocation a greedy sequencing of all tasks in the task graph $G(V,E)$ was performed where the tasks were assigned a weight according to Equation 5, where : $G_v$ is the subgraph rooted at node v and $MeanI(G_v)$ is the mean current of all nodes in the subgraph rooted at node v. Whenever a node is to be scheduled the node with the largest weight was selected among the nodes in the ready list.

$$\forall v \in G \quad w(v) = max\{I_v, MeanI(G_v)\} \quad (5)$$

We executed the two algorithms for three different deadlines for the two different test task graphs G2 and G3 discussed earlier. We present the data for this comparison in Table 4. Notice that as the deadline increases the amount of battery capacity used decreases. This is because the algorithm can choose design points which have lower performance but also consume less capacity. Also, we see that our algorithm gives better results for the two task graphs under consideration.

**Table 4. : Comparison of our algorithm with an approach in [1]**

| | G2: 9 Nodes, 4DPs | | | G3: 15 Nodes, 5 DPs | | |
|---|---|---|---|---|---|---|
| Deadline (*minutes*) | 55 | 75 | 95 | 100 | 150 | 230 |
| Batt. Capacity by Our Algo (*mAmin*) | 30913 | 13751 | 7961 | 57429 | 41801 | 13737 |
| Batt. Capacity by Algo [1] (*mAmin*) | 35739 | 13885 | 8517 | 68120 | 48650 | 22686 |
| % Diff | 15.6 | 0.9 | 7.0 | 18.6 | 16.4 | 65.0 |